\documentclass[aps,prl,superscriptaddress,reprint,floatfix]{revtex4-1}
\usepackage{graphicx}
\usepackage{dcolumn}
\usepackage{bm}
\usepackage[colorlinks,linkcolor=blue,citecolor=blue,urlcolor=blue]{hyperref}
\usepackage{amsmath,amsthm,amssymb}
\usepackage{amsfonts}
\usepackage{icomma}
\usepackage{epstopdf}
\usepackage{placeins}
\usepackage{color}
\usepackage{scalerel}
\usepackage[euler]{textgreek}
\usepackage[separate-uncertainty = true, exponent-product = \times]{siunitx}
\DeclareSIUnit\torr{Torr}
\usepackage[version=4]{mhchem}

\newcommand{\BiSbCo}{$\text{Bi}_{0.74}\text{Sb}_{0.26}$}
\newcommand{\BiSbAl}{$\text{Bi}_{0.96}\text{Sb}_{0.04}$}

\makeatletter
\let\cat@comma@active\@empty
\makeatother

\begin{document}
 
\title{Spin-Orbit Torque and Nernst Effect in Bi-Sb/Co Heterostructures}

\author{Niklas Roschewsky}
\email{roschewsky@berkeley.edu}
\affiliation{Department of Physics, University of California, Berkeley, California 94720, USA}

\author{Emily S. Walker}
\affiliation{Microelectronics Research Center and Department of Electrical and Computer Engineering, University of Texas, Austin, Texas 78758, USA}

\author{Praveen Gowtham}
\affiliation{Department of Electrical Engineering and Computer Sciences, University of California, Berkeley, California 94720, USA}

\author{Sarah Muschinske}
\affiliation{Microelectronics Research Center and Department of Electrical and Computer Engineering, University of Texas, Austin, Texas 78758, USA}

\author{Frances Hellman}
\affiliation{Department of Physics, University of California, Berkeley, California 94720, USA}
\affiliation{Materials Science Division, Lawrence Berkeley National Laboratory, Berkeley, California 94720, USA}

\author{Seth R. Bank}
\affiliation{Microelectronics Research Center and Department of Electrical and Computer Engineering, University of Texas, Austin, Texas 78758, USA}

\author{Sayeef Salahuddin}
\email{sayeef@berkeley.edu}
\affiliation{Department of Electrical Engineering and Computer Sciences, University of California, Berkeley, California 94720, USA}
\affiliation{Materials Science Division, Lawrence Berkeley National Laboratory, Berkeley, California 94720, USA}

\date{\today}

\begin{abstract}
Harmonic measurements of the longitudinal and transverse voltages in Bi-Sb/Co bilayers are presented. A large second harmonic voltage signal due to the ordinary Nernst effect is observed. In experiments where a magnetic field is rotated in the film plane, the ordinary Nernst effect shows the same angular dependence in the transverse voltage as the damping-like spin-orbit torque and in the longitudinal voltage as the unidirectional spin-Hall magneto-resistance respectively. Therefore, the ordinary Nernst effect can be a spurious signal in spin-orbit torque measurements, leading to an overestimation of the spin-Hall angle in topological insulators or semimetals.
\end{abstract}

\maketitle
In metal-based spintronics, heavy metals such as platinum, tantalum or tungsten are used to convert charge currents into spin-currents via the spin-Hall or Rashba-Edelstein effect~\cite{Miron2011,Garello2013,Liu2012a,Liu2012b}. These spin currents can be used to manipulate the magnetization of a magnet, which has useful application in memory technology~\cite{Lee2016,Cubukcu2018}.

One way to quantify the spin-to-charge conversion efficiency is to measure the the spin-Hall angle $\theta_\text{SH}$. While the reported magnitude of the spin-Hall angle varies somewhat in the literature, it is generally agreed that the spin-Hall angle in heavy metals is of the order of $10\%$~\cite{Rojas-Sanchez2014}. To make a significant impact in memory applications, a larger spin-to-charge conversion efficiency, i.e., a larger spin Hall angle is desired~\cite{Apalkov2016,Diao2005}. In this regard, topological insulators gained immense interest recently due to the unique spin-momentum locking they offer, which could lead to significantly higher spin to charge conversion efficiency~\cite{Hellman2017}. Indeed, various research groups have reported the observation of large spin Hall angles compared to those observed in heavy metal layers using various characterization methods, such as harmonic Hall voltage measurements~\cite{Fan2014,DC2018,Li2018}, spin pumping~\cite{Deorani2014,Shiomi2014,Jamali2015,Wang2016,Gupta2018}, spin-Seebeck effect measurements~\cite{Mendes2017}, spin-polarized tunneling studies~\cite{Liu2015,Lee2015}, domain wall motion experiments~\cite{Khang2018,Han2017} and spin-torque ferromagnetic resonance~\cite{Wang2015,Shi2018,Kondou2016,Wang2017,Mellnik2014}. 

In addition, it has been reported that the unidirectional spin-Hall magneto-resistance (USMR) is orders of magnitudes larger in topological insulator/ heavy metal bilayers films than in conventional heavy metal/ ferromagnet bilayers~\cite{Yasuda2016,Ideue2017,Lv2018,He2018}. Since the unidirectional magneto-resistance is asymmetric in the current direction as well as the magnetic field direction, harmonic voltage measurements are often used to quantify this phenomenon~\cite{Avci2015}.

\begin{figure}[b]
    \includegraphics[width=\linewidth]{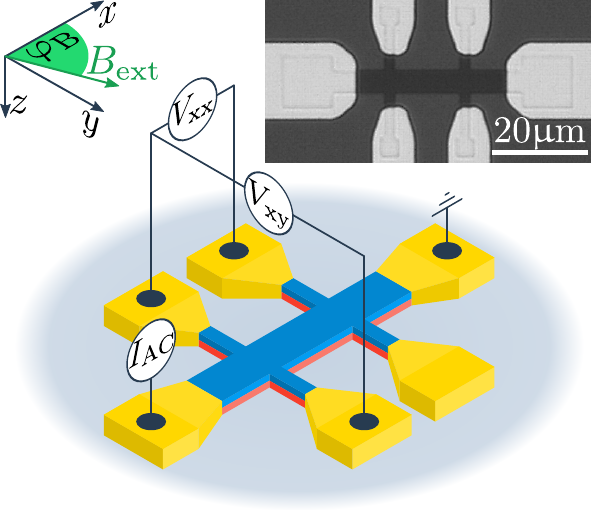}
\caption{\label{fig:Fig1} A double Hall-bar structure is used to measure transverse ($V_{xy}$) and longitudinal ($V_{xx}$) voltages on the Bi-Sb bilayer samples considered in this work. The magnetic field is applied in the film plane at an angle $\varphi$ with respect to the current direction. The inset shows a micro-graph of an actual double Hall-bar device. The width of the Hall bar is \SI{6}{\micro\meter}.}
\end{figure}
While these aforementioned reports have given credence to the promise of using topological insulator materials for high efficiency electronics applications, there are also recent reports that indicate how the extracted spin Hall angles could be impacted by spurious signals. For example, Yasuda \textit{et al.}~\cite{Yasuda2017} have discussed how asymmetric magnon scattering could influence the magnitude of the spin Hall angle. Similarly, it has been observed recently that the Seebeck effect can be a spurious signal in spin-pumping experiments~\cite{Wang2018}. From both the points of view of understanding the underlying physics and its ultimate adoption for an application, it is important to identify the sources of spurious signals in the quantification of the spin Hall angle. In this work, we show that there is an additional source, stemming from the ordinary Nernst effect (ONE)~\cite{Li1970,Yue2018}, that could significantly influence the observations made in harmonic Hall measurements, especially for semiconducting spin Hall materials such as the topological insulators.

Notably, ONE refers to the generation of a voltage, when both a temperature gradient and a magnetic field are present~\cite{Aono1970}. The thermal voltage scales linearly with the external magnetic field and has not been taken into account in previous SOT experiments. Therefore the presence of the ONE in topological insulators at room temperature might give an explanation for the giant spin Hall angles reported recently. Further the ONE can contribute to the longitudinal voltage signal in USMR-type measurements.

Two control experiments will be discussed in the following to separate thermal effects from SOT and USMR. First, we will compare the harmonic voltage response in Bi-Sb/Co samples to that in Bi-Sb/Al samples. While in Bi-Sb/Co, voltages can arise due to magneto-transport effects, the effects in Bi-Sb/Al will be purely thermal. Secondly we will explore the magnetic field dependence of the voltage response. While spin-transport effects are expected to be suppressed by large magnetic fields, the thermal voltage due to the ONE increases linearly with magnetic field.

The samples discussed in this work are \BiSbCo{}(4)/Co(4), \BiSbAl{}(10)/Al(5) and Pt(5)/Py(5) (thickness in \si{\nano\meter}). We refer to these samples as Bi-Sb/Co, Bi-Sb/Al and Pt/Py respectively.  All samples are capped with \SI{2.5}{\nano\meter} $\text{Si}_3\text{N}_4$. The B-Sb is grown epitaxially on high resistivity Si(111) substrate by molecular beam epitaxy. Bulk samples of both \BiSbCo{} and \BiSbAl{} are semi-metals~\cite{Lenoir1996,Jain1959}. It has been reported that a band gap opens up in thin films of \BiSbCo{} due to quantum confinement effects. Therefore \BiSbCo{} films may show topological insulating behavior~\cite{Ueda2017}.

Co is deposited with e-beam evaporation onto the Bi-Sb ex-situ and the magnetic easy axis is in the film plane. Al, Pt and Py are grown by magnetron sputtering. After growth, optical lithography and Ar-ion milling are used to pattern Hall bar devices, as shown in the inset of Fig.~\ref{fig:Fig1}. The width of the Hall bar is \SI{6}{\micro\meter} and the distance between 2 Hall crosses is \SI{20}{\micro\meter}. The current is applied along the $x$-direction while the longitudinal and transverse voltages ($V_{xx}$ and $V_{xy}$) are measured. An external magnetic field $B_\text{ext}$ is applied in the film plane at an angle $\varphi_B$ with respect to the current direction.

In all experiments reported here the magnetic field is sufficient to fully saturate the magnet along the field-direction. Further, all experiments were performed at room temperature and at constant Joule-heating power of $P_\text{heat}=I^2R=\SI{0.18}{\milli\watt}$.

For harmonic measurements presented in this work, an ac current $I=I_0\cdot\sin(\omega t)$ with $\omega/2\pi=\SI{1.5}{\kilo\hertz}$ is applied. If SOT is present, the ac current can induce quasi static magnetization oscillations and thus periodic changes of the resistance and Hall resistance. Therefore, in addition to the first harmonic voltage response $V^\omega$, a second harmonic voltage $V^{2\omega}$ will be induced at twice the excitation frequency~\cite{Avci2014}.

Further, the ac current induces a periodic temperature gradient. Since Joule heating is quadratic in the current, a second harmonic voltage response can also be induced by thermal effects such as the ordinary Nernst effect~\cite{Callen1948,Li1970}, the anomalous Nernst effect~\cite{Chuang2017,Avci2014} or the longitudinal spin-Seebeck effect~\cite{Schreier2013}.

The first harmonic Hall voltage can be written as:
\begin{align*}
  V_{1\omega}^{xy} &= V_\text{PHE}\cdot\sin(2\varphi_B),\\
  V_{1\omega}^{xx} &= V_\text{AMR}\cdot\cos(2\varphi_B).
\end{align*}
Here, $\varphi_B$ is the angle between the current and the external magnetic field $B_\text{ext}$ in the film plane (cf. Fig.~\ref{fig:Fig1}), $V_\text{PHE}$ is the planar Hall effect amplitude and $V_\text{AMR}$ is the amplitude of the anisotropic magneto-resistance. The second harmonic voltage is written as~\cite{Avci2014}:
\begin{widetext}
\begin{alignat}{2}
       V_{xy}^{2\omega}&
      =\bigg(\frac{V_\text{AHE}}{2}\frac{\Delta B_{\text{DL}}}{B_\text{ext}+\mu_0 M_\text{eff}}
            +A \alpha \nabla T
            +N \alpha B_\text{ext} \nabla T \bigg)\,
                 \cos(\varphi_\text{B})
    &+&\bigg(V_\text{PHE}\frac{\Delta B_{\text{FL}}}{B_\text{ext}}\bigg)
                 \cos(2\varphi_\text{B})\cos(\varphi_\text{B})\;,   \label{eq:SOT}\\
       V_{xx}^{2\omega}&
       =\bigg(V_\text{USMR}
            +A \beta \nabla T
            +N \beta B_\text{ext} \nabla T \bigg)\,
                 \sin(\varphi_\text{B})
    &+&\bigg(V_\text{AMR}\frac{\Delta B_{\text{FL}}}{B_\text{ext}}\bigg)
                 \sin(2\varphi_\text{B})\cos(\varphi_\text{B})\;,   \label{eq:USMR}
\end{alignat}
\end{widetext}
where $V_\text{AHE}$ is the anomalous Hall effect voltage, $\Delta B_\text{DL}$ is the effective field due to the damping like torque, $\mu_0 M_\text{eff}$ is the effective magnetization, A is the coefficient for the ANE/ SSE, $\alpha$ and $\beta$ are geometrical factors, $\Delta T$ is the temperature gradient, N is the ordinary Nernst coefficient, $\Delta B_\text{DL}$ is the effective field due to field-like torque and $V_\text{USMR}$ is the amplitude of the unidirectional spin-Hall magneto-resistance.

The field-like SOT can be extracted from the $\varphi_\text{B}$ dependence of the second Harmonic voltage signal. However, it is important to consider the magnetic field dependence in addition to the $\varphi_\text{B}$-dependence for the extraction of the damping-like SOT, as well as the USMR, to distinguish magneto-thermal effects from SOT.

%
\begin{figure}[t]
    \begin{center}
\begingroup
  \makeatletter
  \providecommand\color[2][]{%
    \GenericError{(gnuplot) \space\space\space\@spaces}{%
      Package color not loaded in conjunction with
      terminal option `colourtext'%
    }{See the gnuplot documentation for explanation.%
    }{Either use 'blacktext' in gnuplot or load the package
      color.sty in LaTeX.}%
    \renewcommand\color[2][]{}%
  }%
  \providecommand\includegraphics[2][]{%
    \GenericError{(gnuplot) \space\space\space\@spaces}{%
      Package graphicx or graphics not loaded%
    }{See the gnuplot documentation for explanation.%
    }{The gnuplot epslatex terminal needs graphicx.sty or graphics.sty.}%
    \renewcommand\includegraphics[2][]{}%
  }%
  \providecommand\rotatebox[2]{#2}%
  \@ifundefined{ifGPcolor}{%
    \newif\ifGPcolor
    \GPcolortrue
  }{}%
  \@ifundefined{ifGPblacktext}{%
    \newif\ifGPblacktext
    \GPblacktextfalse
  }{}%
  \let\gplgaddtomacro\g@addto@macro
  \gdef\gplbacktext{}%
  \gdef\gplfronttext{}%
  \makeatother
  \ifGPblacktext
    \def\colorrgb#1{}%
    \def\colorgray#1{}%
  \else
    \ifGPcolor
      \def\colorrgb#1{\color[rgb]{#1}}%
      \def\colorgray#1{\color[gray]{#1}}%
      \expandafter\def\csname LTw\endcsname{\color{white}}%
      \expandafter\def\csname LTb\endcsname{\color{black}}%
      \expandafter\def\csname LTa\endcsname{\color{black}}%
      \expandafter\def\csname LT0\endcsname{\color[rgb]{1,0,0}}%
      \expandafter\def\csname LT1\endcsname{\color[rgb]{0,1,0}}%
      \expandafter\def\csname LT2\endcsname{\color[rgb]{0,0,1}}%
      \expandafter\def\csname LT3\endcsname{\color[rgb]{1,0,1}}%
      \expandafter\def\csname LT4\endcsname{\color[rgb]{0,1,1}}%
      \expandafter\def\csname LT5\endcsname{\color[rgb]{1,1,0}}%
      \expandafter\def\csname LT6\endcsname{\color[rgb]{0,0,0}}%
      \expandafter\def\csname LT7\endcsname{\color[rgb]{1,0.3,0}}%
      \expandafter\def\csname LT8\endcsname{\color[rgb]{0.5,0.5,0.5}}%
    \else
      \def\colorrgb#1{\color{black}}%
      \def\colorgray#1{\color[gray]{#1}}%
      \expandafter\def\csname LTw\endcsname{\color{white}}%
      \expandafter\def\csname LTb\endcsname{\color{black}}%
      \expandafter\def\csname LTa\endcsname{\color{black}}%
      \expandafter\def\csname LT0\endcsname{\color{black}}%
      \expandafter\def\csname LT1\endcsname{\color{black}}%
      \expandafter\def\csname LT2\endcsname{\color{black}}%
      \expandafter\def\csname LT3\endcsname{\color{black}}%
      \expandafter\def\csname LT4\endcsname{\color{black}}%
      \expandafter\def\csname LT5\endcsname{\color{black}}%
      \expandafter\def\csname LT6\endcsname{\color{black}}%
      \expandafter\def\csname LT7\endcsname{\color{black}}%
      \expandafter\def\csname LT8\endcsname{\color{black}}%
    \fi
  \fi
    \setlength{\unitlength}{0.0500bp}%
    \ifx\gptboxheight\undefined%
      \newlength{\gptboxheight}%
      \newlength{\gptboxwidth}%
      \newsavebox{\gptboxtext}%
    \fi%
    \setlength{\fboxrule}{0.5pt}%
    \setlength{\fboxsep}{1pt}%
\begin{picture}(4874.00,3968.00)%
    \gplgaddtomacro\gplbacktext{%
      \csname LTb\endcsname%
      \put(660,2622){\makebox(0,0)[r]{\strut{}$-0.1$}}%
      \put(660,3074){\makebox(0,0)[r]{\strut{}$0$}}%
      \put(660,3526){\makebox(0,0)[r]{\strut{}$0.1$}}%
      \put(792,2041){\makebox(0,0){\strut{}}}%
      \put(1779,2041){\makebox(0,0){\strut{}}}%
      \put(2767,2041){\makebox(0,0){\strut{}}}%
      \put(3754,2041){\makebox(0,0){\strut{}}}%
      \put(4741,2041){\makebox(0,0){\strut{}}}%
      \put(3534,3706){\makebox(0,0)[l]{\strut{}$B_\text{ext}=\SI{90}{\milli\tesla}$}}%
    }%
    \gplgaddtomacro\gplfronttext{%
      \csname LTb\endcsname%
      \put(102,3074){\rotatebox{-270}{\makebox(0,0){\strut{}$V^{xy}_{1\omega} (\si{\milli\volt})$}}}%
      \colorrgb{0.00,0.38,0.68}%
      \put(1549,2829){\makebox(0,0)[r]{\strut{}BiSb/Co}}%
      \colorrgb{0.87,0.09,0.12}%
      \put(1549,2609){\makebox(0,0)[r]{\strut{}BiSb/Al}}%
      \colorrgb{0.38,0.70,0.28}%
      \put(1549,2389){\makebox(0,0)[r]{\strut{}Pt/Py}}%
    }%
    \gplgaddtomacro\gplbacktext{%
      \csname LTb\endcsname%
      \put(660,959){\makebox(0,0)[r]{\strut{}$-1.5$}}%
      \put(660,1448){\makebox(0,0)[r]{\strut{}$0$}}%
      \put(660,1936){\makebox(0,0)[r]{\strut{}$1.5$}}%
      \put(792,414){\makebox(0,0){\strut{}$-180$}}%
      \put(1779,414){\makebox(0,0){\strut{}$-90$}}%
      \put(2767,414){\makebox(0,0){\strut{}$0$}}%
      \put(3754,414){\makebox(0,0){\strut{}$90$}}%
      \put(4741,414){\makebox(0,0){\strut{}$180$}}%
      \put(3315,2098){\makebox(0,0)[l]{\strut{}$P_\text{heat}=\SI{0.18}{\milli\watt}$}}%
    }%
    \gplgaddtomacro\gplfronttext{%
      \csname LTb\endcsname%
      \put(102,1447){\rotatebox{-270}{\makebox(0,0){\strut{}$V^{xy}_{2\omega} (\si{\micro\volt})$}}}%
      \put(2766,84){\makebox(0,0){\strut{}$\varphi_\text{B}$ (deg)}}%
    }%
    \gplbacktext
    \put(0,0){\includegraphics{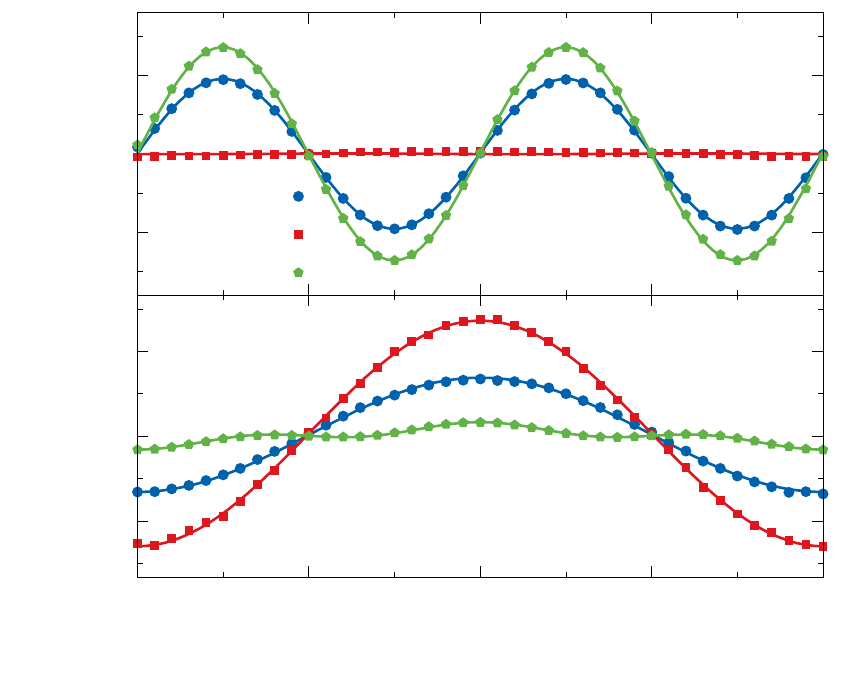}}%
    \gplfronttext
  \end{picture}%
\endgroup
    \end{center}
\caption{\label{fig:Fig2} Upper panel: The planar Hall effect is only observed in the samples with magnetic layers ( Bi-Sb/Co and Pt/Py). Here, $\varphi_\text B$ is the angle between the current direction and $B_\text{ext}$ in the film plane. Lower panel: $V_{2\omega}$ shows a $\cos(\varphi_\text B)$ dependence in the samples with Bi-Sb, due to the ordinary Nernst effect. In the sample Pt/Py, spin-orbit torque and anomalous Nernst effect are present. The solid lines are fits to eq.\eqref{eq:SOT}.}
\end{figure}
We start by discussing the angular dependence of the transverse harmonic voltages. The top panel in Fig.~\ref{fig:Fig2} shows $V^{1\omega}_{xy}$ as a function of $\varphi_\text{B}$. In the samples Bi-Sb/Co and Pt/Py, a $\sin(2\varphi_\text{B})$-dependence can be seen. This is due to the presence of the PHE in the Co and Py layers respectively. $V^{1\omega}_{xy}$ does not depend on $B_\text{ext}$ in Bi-Sb/Al, due to the absence of a magnetic layer.

The second harmonic Hall voltage $V_{xy}^{2\omega}$ is shown in the bottom panel of Fig.~\ref{fig:Fig2}. In all 3 samples under consideration, $V_{xy}^{2\omega}$ vs. $\varphi_\text{B}$ can be well fitted by Eq.~\eqref{eq:SOT} (solid lines). In the Pt/Py sample contributions from the field-like torque are clearly visible. On the other hand, samples Bi-Sb/Co and Bi-Sb/Al do not show the characteristic $\cos(2\varphi_\text{B})\cos(\varphi_\text{B})$-dependence. Since no magnet is present in the sample Bi-Sb/Al, the large second harmonic voltage is attributed to the ONE.

The fact that no field-like torque is observed in the sample Bi-Sb/Co, as well as the fact that voltage signals in Bi-Sb/Al and Bi-Sb/Co are comparable gives rise to the suspicion that the signal in the sample Bi-Sb/Co might be dominated by the ONE. To unambiguously distinguish the ONE from SOT, the $B_\text{ext}$-dependence will have to be considered, which we will discuss after the following discussion on USMR measurements.

\begin{figure}[htp]
    \begin{center}
\begingroup
  \makeatletter
  \providecommand\color[2][]{%
    \GenericError{(gnuplot) \space\space\space\@spaces}{%
      Package color not loaded in conjunction with
      terminal option `colourtext'%
    }{See the gnuplot documentation for explanation.%
    }{Either use 'blacktext' in gnuplot or load the package
      color.sty in LaTeX.}%
    \renewcommand\color[2][]{}%
  }%
  \providecommand\includegraphics[2][]{%
    \GenericError{(gnuplot) \space\space\space\@spaces}{%
      Package graphicx or graphics not loaded%
    }{See the gnuplot documentation for explanation.%
    }{The gnuplot epslatex terminal needs graphicx.sty or graphics.sty.}%
    \renewcommand\includegraphics[2][]{}%
  }%
  \providecommand\rotatebox[2]{#2}%
  \@ifundefined{ifGPcolor}{%
    \newif\ifGPcolor
    \GPcolortrue
  }{}%
  \@ifundefined{ifGPblacktext}{%
    \newif\ifGPblacktext
    \GPblacktextfalse
  }{}%
  \let\gplgaddtomacro\g@addto@macro
  \gdef\gplbacktext{}%
  \gdef\gplfronttext{}%
  \makeatother
  \ifGPblacktext
    \def\colorrgb#1{}%
    \def\colorgray#1{}%
  \else
    \ifGPcolor
      \def\colorrgb#1{\color[rgb]{#1}}%
      \def\colorgray#1{\color[gray]{#1}}%
      \expandafter\def\csname LTw\endcsname{\color{white}}%
      \expandafter\def\csname LTb\endcsname{\color{black}}%
      \expandafter\def\csname LTa\endcsname{\color{black}}%
      \expandafter\def\csname LT0\endcsname{\color[rgb]{1,0,0}}%
      \expandafter\def\csname LT1\endcsname{\color[rgb]{0,1,0}}%
      \expandafter\def\csname LT2\endcsname{\color[rgb]{0,0,1}}%
      \expandafter\def\csname LT3\endcsname{\color[rgb]{1,0,1}}%
      \expandafter\def\csname LT4\endcsname{\color[rgb]{0,1,1}}%
      \expandafter\def\csname LT5\endcsname{\color[rgb]{1,1,0}}%
      \expandafter\def\csname LT6\endcsname{\color[rgb]{0,0,0}}%
      \expandafter\def\csname LT7\endcsname{\color[rgb]{1,0.3,0}}%
      \expandafter\def\csname LT8\endcsname{\color[rgb]{0.5,0.5,0.5}}%
    \else
      \def\colorrgb#1{\color{black}}%
      \def\colorgray#1{\color[gray]{#1}}%
      \expandafter\def\csname LTw\endcsname{\color{white}}%
      \expandafter\def\csname LTb\endcsname{\color{black}}%
      \expandafter\def\csname LTa\endcsname{\color{black}}%
      \expandafter\def\csname LT0\endcsname{\color{black}}%
      \expandafter\def\csname LT1\endcsname{\color{black}}%
      \expandafter\def\csname LT2\endcsname{\color{black}}%
      \expandafter\def\csname LT3\endcsname{\color{black}}%
      \expandafter\def\csname LT4\endcsname{\color{black}}%
      \expandafter\def\csname LT5\endcsname{\color{black}}%
      \expandafter\def\csname LT6\endcsname{\color{black}}%
      \expandafter\def\csname LT7\endcsname{\color{black}}%
      \expandafter\def\csname LT8\endcsname{\color{black}}%
    \fi
  \fi
    \setlength{\unitlength}{0.0500bp}%
    \ifx\gptboxheight\undefined%
      \newlength{\gptboxheight}%
      \newlength{\gptboxwidth}%
      \newsavebox{\gptboxtext}%
    \fi%
    \setlength{\fboxrule}{0.5pt}%
    \setlength{\fboxsep}{1pt}%
\begin{picture}(4874.00,3968.00)%
    \gplgaddtomacro\gplbacktext{%
      \csname LTb\endcsname%
      \put(660,2460){\makebox(0,0)[r]{\strut{}$-0.4$}}%
      \put(660,3074){\makebox(0,0)[r]{\strut{}$0$}}%
      \put(660,3688){\makebox(0,0)[r]{\strut{}$0.4$}}%
      \put(792,2041){\makebox(0,0){\strut{}}}%
      \put(1779,2041){\makebox(0,0){\strut{}}}%
      \put(2767,2041){\makebox(0,0){\strut{}}}%
      \put(3754,2041){\makebox(0,0){\strut{}}}%
      \put(4741,2041){\makebox(0,0){\strut{}}}%
      \put(3480,3718){\makebox(0,0)[l]{\strut{}$B_\text{ext}=\SI{90}{\milli\tesla}$}}%
    }%
    \gplgaddtomacro\gplfronttext{%
      \csname LTb\endcsname%
      \put(102,3074){\rotatebox{-270}{\makebox(0,0){\strut{}$V^{xx}_{1\omega} (\si{\milli\volt})$}}}%
    }%
    \gplgaddtomacro\gplbacktext{%
      \csname LTb\endcsname%
      \put(660,837){\makebox(0,0)[r]{\strut{}$-6$}}%
      \put(660,1448){\makebox(0,0)[r]{\strut{}$0$}}%
      \put(660,2058){\makebox(0,0)[r]{\strut{}$6$}}%
      \put(792,414){\makebox(0,0){\strut{}$-180$}}%
      \put(1779,414){\makebox(0,0){\strut{}$-90$}}%
      \put(2767,414){\makebox(0,0){\strut{}$0$}}%
      \put(3754,414){\makebox(0,0){\strut{}$90$}}%
      \put(4741,414){\makebox(0,0){\strut{}$180$}}%
      \put(3227,807){\makebox(0,0)[l]{\strut{}$P_\text{heat}=\SI{0.18}{\milli\watt}$}}%
    }%
    \gplgaddtomacro\gplfronttext{%
      \csname LTb\endcsname%
      \put(102,1447){\rotatebox{-270}{\makebox(0,0){\strut{}$V^{xx}_{2\omega} (\si{\micro\volt})$}}}%
      \put(2766,84){\makebox(0,0){\strut{}$\varphi_\text{B}$ (deg)}}%
      \colorrgb{0.00,0.38,0.68}%
      \put(1604,2090){\makebox(0,0)[r]{\strut{}BiSb/Co}}%
      \colorrgb{0.87,0.09,0.12}%
      \put(1604,1870){\makebox(0,0)[r]{\strut{}BiSb/Al}}%
      \colorrgb{0.38,0.70,0.28}%
      \put(1604,1650){\makebox(0,0)[r]{\strut{}Pt/Py}}%
    }%
    \gplbacktext
    \put(0,0){\includegraphics{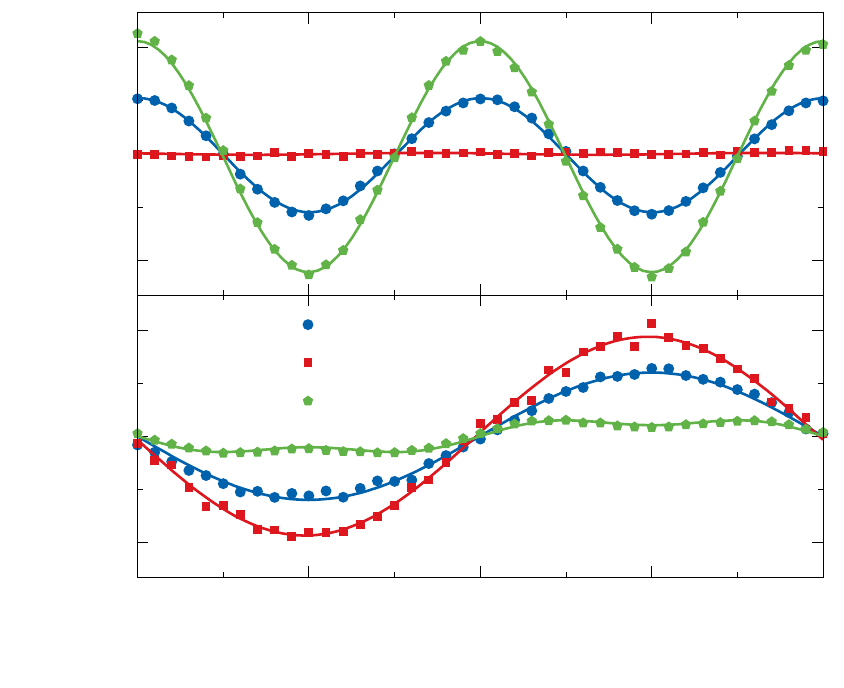}}%
    \gplfronttext
  \end{picture}%
\endgroup
    \end{center}
\caption{\label{fig:Fig3} Upper panel: $V_{1\omega}^{xx}$ as a function of $\varphi_\text B$. A constant offset was removed for easy comparison. The anisotropic magneto-resistance is observed only in the magnetic samples, Bi-Sb/Co and Pt/Py. Lower panel: $V_{2\omega}^{xx}$ in Bi-Sb/Co and Bi-Sb/Al shows a $\sin(\varphi)$ dependence. This is expected from the ordinary Nernst effect. In the sample Pt/Py  contributions from spin-orbit torque, anomalous Nernst effect and the unidirectional spin-Hall magneto-resistance are observed. The solid lines are fits to Eq.\eqref{eq:USMR}.}
\end{figure}
In addition to the angular dependence of $V_{xy}$, the longitudinal voltage $V_{xx}$ was measured (Fig.~\ref{fig:Fig3}). Again, a signal from the anisotropic magneto-resistance (AMR) is only observed in the samples with a magnetic layer (top panel), while all samples show an angular dependence of $V_{2\omega}^{xx}$ that can be well described with Eq.~\eqref{eq:USMR} (bottom panel). In the Pt/Py reference sample, a clear signature from field-like SOT ($\propto\sin(2\varphi_\text{B})\cos(\varphi_\text{B})$) can be observed while the USMR ($\propto\sin(\varphi_\text{B})$) is small. The sample Bi-Sb/Al shows a $\sin(\varphi)$-dependence due to the ONE. The absence of field-like torque in Bi-Sb/Co, as well as a signal amplitude comparable to the signal in Bi-Sb/Al raises the suspicion that the signal in Bi-Sb/Co is dominated by the ONE and not a large USMR.

\begin{figure}[t]
  \begin{center}
\begingroup
  \makeatletter
  \providecommand\color[2][]{%
    \GenericError{(gnuplot) \space\space\space\@spaces}{%
      Package color not loaded in conjunction with
      terminal option `colourtext'%
    }{See the gnuplot documentation for explanation.%
    }{Either use 'blacktext' in gnuplot or load the package
      color.sty in LaTeX.}%
    \renewcommand\color[2][]{}%
  }%
  \providecommand\includegraphics[2][]{%
    \GenericError{(gnuplot) \space\space\space\@spaces}{%
      Package graphicx or graphics not loaded%
    }{See the gnuplot documentation for explanation.%
    }{The gnuplot epslatex terminal needs graphicx.sty or graphics.sty.}%
    \renewcommand\includegraphics[2][]{}%
  }%
  \providecommand\rotatebox[2]{#2}%
  \@ifundefined{ifGPcolor}{%
    \newif\ifGPcolor
    \GPcolortrue
  }{}%
  \@ifundefined{ifGPblacktext}{%
    \newif\ifGPblacktext
    \GPblacktextfalse
  }{}%
  \let\gplgaddtomacro\g@addto@macro
  \gdef\gplbacktext{}%
  \gdef\gplfronttext{}%
  \makeatother
  \ifGPblacktext
    \def\colorrgb#1{}%
    \def\colorgray#1{}%
  \else
    \ifGPcolor
      \def\colorrgb#1{\color[rgb]{#1}}%
      \def\colorgray#1{\color[gray]{#1}}%
      \expandafter\def\csname LTw\endcsname{\color{white}}%
      \expandafter\def\csname LTb\endcsname{\color{black}}%
      \expandafter\def\csname LTa\endcsname{\color{black}}%
      \expandafter\def\csname LT0\endcsname{\color[rgb]{1,0,0}}%
      \expandafter\def\csname LT1\endcsname{\color[rgb]{0,1,0}}%
      \expandafter\def\csname LT2\endcsname{\color[rgb]{0,0,1}}%
      \expandafter\def\csname LT3\endcsname{\color[rgb]{1,0,1}}%
      \expandafter\def\csname LT4\endcsname{\color[rgb]{0,1,1}}%
      \expandafter\def\csname LT5\endcsname{\color[rgb]{1,1,0}}%
      \expandafter\def\csname LT6\endcsname{\color[rgb]{0,0,0}}%
      \expandafter\def\csname LT7\endcsname{\color[rgb]{1,0.3,0}}%
      \expandafter\def\csname LT8\endcsname{\color[rgb]{0.5,0.5,0.5}}%
    \else
      \def\colorrgb#1{\color{black}}%
      \def\colorgray#1{\color[gray]{#1}}%
      \expandafter\def\csname LTw\endcsname{\color{white}}%
      \expandafter\def\csname LTb\endcsname{\color{black}}%
      \expandafter\def\csname LTa\endcsname{\color{black}}%
      \expandafter\def\csname LT0\endcsname{\color{black}}%
      \expandafter\def\csname LT1\endcsname{\color{black}}%
      \expandafter\def\csname LT2\endcsname{\color{black}}%
      \expandafter\def\csname LT3\endcsname{\color{black}}%
      \expandafter\def\csname LT4\endcsname{\color{black}}%
      \expandafter\def\csname LT5\endcsname{\color{black}}%
      \expandafter\def\csname LT6\endcsname{\color{black}}%
      \expandafter\def\csname LT7\endcsname{\color{black}}%
      \expandafter\def\csname LT8\endcsname{\color{black}}%
    \fi
  \fi
    \setlength{\unitlength}{0.0500bp}%
    \ifx\gptboxheight\undefined%
      \newlength{\gptboxheight}%
      \newlength{\gptboxwidth}%
      \newsavebox{\gptboxtext}%
    \fi%
    \setlength{\fboxrule}{0.5pt}%
    \setlength{\fboxsep}{1pt}%
\begin{picture}(4874.00,3968.00)%
    \gplgaddtomacro\gplbacktext{%
      \csname LTb\endcsname%
      \put(660,2261){\makebox(0,0)[r]{\strut{}$0$}}%
      \put(660,2803){\makebox(0,0)[r]{\strut{}$3$}}%
      \put(660,3345){\makebox(0,0)[r]{\strut{}$6$}}%
      \put(660,3887){\makebox(0,0)[r]{\strut{}$9$}}%
      \put(792,2041){\makebox(0,0){\strut{}}}%
      \put(2108,2041){\makebox(0,0){\strut{}}}%
      \put(3425,2041){\makebox(0,0){\strut{}}}%
      \put(4741,2041){\makebox(0,0){\strut{}}}%
      \colorrgb{0.38,0.70,0.28}%
      \put(1055,2803){\rotatebox{-2}{\makebox(0,0)[l]{\strut{}$50\cdot x$}}}%
    }%
    \gplgaddtomacro\gplfronttext{%
      \csname LTb\endcsname%
      \put(286,3074){\rotatebox{-270}{\makebox(0,0){\strut{}$V^{xy}_{2\omega} (\si{\micro\volt})$}}}%
      \put(2766,1975){\makebox(0,0){\strut{}}}%
      \colorrgb{0.00,0.38,0.68}%
      \put(1637,3687){\makebox(0,0)[r]{\strut{}BiSb/Co}}%
      \colorrgb{0.87,0.09,0.12}%
      \put(1637,3467){\makebox(0,0)[r]{\strut{}BiSb/Al}}%
      \colorrgb{0.38,0.70,0.28}%
      \put(1637,3247){\makebox(0,0)[r]{\strut{}Pt/Py}}%
    }%
    \gplgaddtomacro\gplbacktext{%
      \csname LTb\endcsname%
      \put(660,634){\makebox(0,0)[r]{\strut{}0}}%
      \put(660,1041){\makebox(0,0)[r]{\strut{}5}}%
      \put(660,1448){\makebox(0,0)[r]{\strut{}10}}%
      \put(660,1854){\makebox(0,0)[r]{\strut{}15}}%
      \put(660,2261){\makebox(0,0)[r]{\strut{}}}%
      \put(792,414){\makebox(0,0){\strut{}$0$}}%
      \put(2108,414){\makebox(0,0){\strut{}$100$}}%
      \put(3425,414){\makebox(0,0){\strut{}$200$}}%
      \put(4741,414){\makebox(0,0){\strut{}$300$}}%
      \put(3227,878){\makebox(0,0)[l]{\strut{}$P_\text{heat}=\SI{0.18}{\milli\watt}$}}%
      \colorrgb{0.38,0.70,0.28}%
      \put(1055,1366){\rotatebox{-2}{\makebox(0,0)[l]{\strut{}$10\cdot x$}}}%
    }%
    \gplgaddtomacro\gplfronttext{%
      \csname LTb\endcsname%
      \put(286,1447){\rotatebox{-270}{\makebox(0,0){\strut{}$V^{xx}_{2\omega} (\si{\micro\volt})$}}}%
      \put(2766,84){\makebox(0,0){\strut{}$B_\text{ext}$ (mT)}}%
    }%
    \gplbacktext
    \put(0,0){\includegraphics{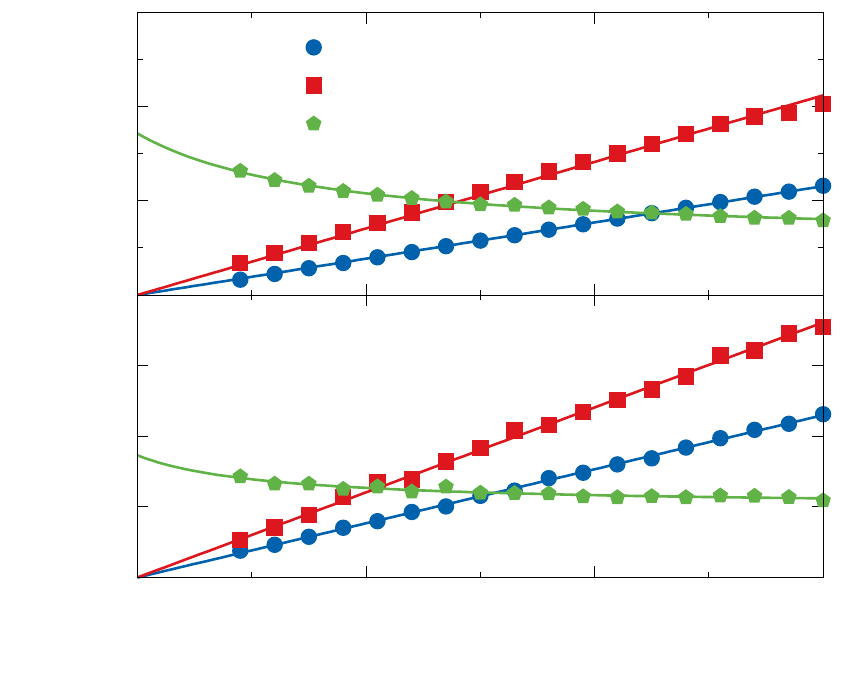}}%
    \gplfronttext
  \end{picture}%
\endgroup
    \caption{\label{fig:Fig4} The magnetic field dependence of the second harmonic voltage. In Bi-Sb/Co and Bi-Sb/Al a linear increase in voltage can be observed, as expected from the ordinary Nernst effect. In Pt/Py on the other hand the harmonic voltage is decreasing as expected from damping-like spin-orbit torque and unidirectional spin-Hall magneto-resistance respectively. Note that the voltage in Pt/Py has been enhances for visual clarity.}
  \end{center}
\end{figure}

To determine the origin of the large voltages observed in Bi-Sb/Co, we repeat the measurements shown in Fig.~\ref{fig:Fig2} and Fig.~\ref{fig:Fig3} for different values of $B_\text{ext}$. Figure~\ref{fig:Fig4} shows the $\cos(\varphi_B)$ contribution to $V^{xy}_{2\omega}$ and the $\sin(\varphi_B)$ contribution to $V^{xx}_{2\omega}$ respectively. In the reference sample Pt/Py, $V^{xy}_{2\omega}$ decreases with increasing $B_\text{ext}$. This is expected, because large magnetic fields suppress the quasi-static magnetization oscillations induced by SOT (cf. Eq.~\eqref{eq:SOT}). On the other hand, $V^{xy}_{2\omega}$ increases linearly with $B_\text{ext}$ in the samples Bi-Sb/Co and Bi-Sb/Al. This is clear evidence for the presence of ONE. Note that $V^{xy}_{2\omega}$ is $\approx 50$ times larger in the samples with Bi-Sb.

Similarly, a clear presence for the USMR is seen only in the sample Pt/Py (Fig.~\ref{fig:Fig4}, bottom panel): $V_{2\omega}^{xx}$ decreases with increasing $B_\text{ext}$, in accordance with recent reports by Avci \textit{et al}.~\cite{Avci2018} The samples Bi-Sb/Al and Bi-Sb/Co show a linear dependence on $B_\text{ext}$, as expected from the ONE. The slope of $V^{xx}_{2\omega}$ vs $B_\text{ext}$ in Bi-Sb/Co is $V^{xy}_{2\omega}/B_\text{ext}=\SI{0.011(6)}{\volt\per\milli\tesla}$ and $V^{xx}_{2\omega}/B_\text{ext}=\SI{0.038(3)}{\volt\per\milli\tesla}$ respectively. The different values obtained for the longitudinal and transverse voltages are explained by the geometric factors $\alpha$ and $\beta$ in eq.~\ref{eq:SOT} and Eq.~\ref{eq:USMR}. The ratio is $V^{xx}_{2\omega}/V^{xy}_{2\omega}=\beta/\alpha=l/w=3.3$, where $l$ is the distance between two Hall crosses and $w$ the width of the Hall bar. This geometrical scaling is a strong indicator that $V_{2\omega}^{xy}$ and $V_{2\omega}^{xx}$ have the same physical origin. Note that USMR and SOT are distinctly different phenomena and it can not be expected that they would scale with the device geometry.

Next we will calculate the ONE coefficient N and show that N is equal in samples Bi-Sb/Co and Bi-Sb/Al. To this end we estimate the temperature gradient using the 1D heat equation $T''(z)=\dot q_i/\kappa$, where $T(z)$ is the temperature, $\dot q_i$ is the volumetric power density and $\kappa$ the thermal conductivity. The boundary conditions at the interfaces are:
\begin{align*}
  T_i=T_{i+1} \;,\;\;\;\; \kappa_i\frac{\partial T_i}{\partial z}=\kappa_{i+1}\frac{\partial T_{i+1}}{\partial z}
\end{align*}
We assume no heat conduction at the surface of the sample and a fixed temperature at the backside of the substrate. The following values for the thermal conductivity have been used: $\kappa_\text{Bi-Sb}=\SI{6.8}{\watt\per\meter\per\kelvin}$~\cite{Yazaki1968}, $\kappa_\text{Co}=\SI{98}{\watt\per\meter\per\kelvin}$~\cite{Terada2003} and $\kappa_\text{Al}=\SI{238}{\watt\per\meter\per\kelvin}$~\cite{Paul1993}. At a heating power of $P=\SI{180}{\micro\watt}$, the temperature gradient is $\Delta T_\text{Bi-Sb/Co}=\SI{6.6}{\milli\kelvin}$ and $\Delta T_\text{Bi-Sb/Al}=\SI{18}{\milli\kelvin}$. Using these values, we find $N=\SI{3e-6}{\volt\per\tesla\per\kelvin}$ in Bi-Sb/Co and $N=\SI{2.6e-6}{\volt\per\tesla\per\kelvin}$ in Bi-Sb/Al. These values are well in agreement and confirm the same origin of $V_{2\omega}$ in the samples Bi-Sb/Co and Bi-Sb/Al, namely the ordinary Nernst effect. Further, $N$ is of the same order of magnitude in other semi-conducting materials~\cite{Behnia2007}.

In summary, harmonic measurements of the longitudinal and transverse voltages in Bi-Sb/Co films are presented. We find strong evidence that the second harmonic voltage in our samples is dominated by contributions from the ordinary Nernst effect: The longitudinal and transverse voltages  scale with the device geometry. Further, we show that the ordinary Nernst effect has the same amplitude in Bi-Sb/Al samples where magneto-transport effects are absent. In fact, we find that the voltages from the ordinary Nernst effect is an order of magnitude larger than voltages expected from SOT or USMR, which makes the detection of pure spin related effects highly challenging. The low thermal conductivity of Bi-Sb due to it's semiconducting nature is predominantly responsible for the large ONE related voltages observed in our experiments. Given that the symmetry of the ONE related voltages are the same as those originating from spin orbit torque, it is critical to measure the harmonic voltages as a function of the external magnetic field amplitude to distinguish different contributions such as ANE, ONE and SOT/USMR.

\vspace{0.2 cm}
The authors thank Johannes Mendil and Dominic Labanowski for fruitful discussions. Research was supported by the U.S. Department of Energy under Contract No. DE-AC02-05-CH11231 within the NEMM program (KC2204) and NSF Materials Research Science and Engineering Center Grant No. DMR-1720595.  Device fabrication was supported by NSF E3S center at Berkeley and STARNET/LEAST, one of the six centers of the JUMP initiative jointly supported by DARPA/SRC.


%

\end{document}